\documentclass[tradiabstract]{aa}
\usepackage{txfonts,epsfig,graphicx,natbib,url,twoopt}
\usepackage{amssymb}
\usepackage{aalongtable}
\usepackage{longtable}
\usepackage{indentfirst}
\usepackage{natbib} 
\bibpunct{(}{)}{;}{a}{}{,} % to follow the A&A style
\newcommand{\ie}{{\em i.e. }} 
\newcommand{\eg}{{\em e.g. }}
\newcommand{\vs}{{\em vs. }}

\usepackage[breaklinks=true]{hyperref} %% to avoid \citeads line fills
\usepackage[labelfont=footnotesize,textfont=footnotesize]{caption}  %% as A&A

\bibpunct{(}{)}{;}{a}{}{,}    %% natbib format choice of A&A and ApJ
\newcommandtwoopt{\citeads}[3][][]{\href{http://adsabs.harvard.edu/abs/#3}%
                                        {\citealp[#1][#2]{#3}}}
\newcommandtwoopt{\citepads}[3][][]{\href{http://adsabs.harvard.edu/abs/#3}%
                                         {\citep[#1][#2]{#3}}}
\newcommandtwoopt{\citetads}[3][][]{\href{http://adsabs.harvard.edu/abs/#3}%
                                         {\citet[#1][#2]{#3}}}
\newcommandtwoopt{\citeyearrads}%
  [3][][]{\href{http://adsabs.harvard.edu/abs/#3}%
                                         {\citeyear[#1][#2]{#3}}}

\begin{document}  

\twocolumn[{%
\vspace*{4ex}
\begin{center}
  {\Large \bf The Bimodal Colors of Centaurs and Small Kuiper Belt Objects}\\[4ex]  
  {\large \bf Nuno Peixinho$^{1, 2}$, 
              Audrey Delsanti$^{3,4}$,
              Aur\'elie Guilbert-Lepoutre$^5$,
              Ricardo Gafeira$^1$,
              and Pedro Lacerda$^6$}\\[4ex]
  \begin{minipage}[t]{15cm}
        $^1$ Center for Geophysics of the University of Coimbra, Av. Dr. Dias da Silva, 3000-134 Coimbra,  Portugal 
        [{\it e-mail:} peixinho@mat.uc.pt; gafeira@mat.uc.pt]\\
        $^2$ Astronomical Observatory of the University of Coimbra, Almas de Freire, 3040-004 Coimbra, Portugal \\
        $^3$ Laboratoire dÕAstrophysique de Marseille, Universit\'e d'Aix-Marseille, CNRS, 38 rue Fr\'ed\'eric Joliot-Curie, 13388 Marseille, France
        [{\it e-mail:} Audrey.Delsanti@oamp.fr; Audrey.Delsanti@obspm.fr]\\
        $^4$ Observatoire de Paris, Site de Meudon, 5 place Jules Janssen, 92190 Meudon, France\\
        $^5$ UCLA, Department of Earth and Space Sciences, 595 Charles E. Young Drive East, Los Angeles CA 90095, USA 
        [{\it email:} aguilbert@ucla.edu]\\
        $^6$ Queen's University Belfast, Astrophysics Research Centre, Belfast BT7 1NN, United Kingdom
        [{\it email:} p.lacerda@qub.ac.uk]\\
        \\
        {\it To appear in Astronomy \& Astrophysics}\\

%%%%%%%%%%%%%%%%%%%%%%%%%%%%%%%%%%%%%%%%%%%%%%%%%%%%%%%%%%%%%%%%%%%%%%%%%%%%
%%%%%%%%%%%%%%%%%%%%%%%%%%%%%%%%%%%%%%%%%%%%%%%%%%%%%%%%%%%%%%%%%%%%%%%%%%%%

{\bf Abstract.}  Ever since the very first photometric studies of Centaurs and Kuiper Belt Objects (KBOs)
their visible color distribution has been controversial. 
That controversy gave rise to a prolific debate on the origin of the surface colors of these distant icy objects 
of the Solar System. 
Two different views attempt to interpret and explain the large variability of colors, hence surface composition.  
Are the colors mainly primordial and directly related to the formation region, or are they the result of surface evolution processes?
To date, no mechanism has been found that successfully explains why Centaurs, which are escapees from the Kuiper Belt, 
exhibit two distinct color groups, whereas KBOs do not. 
In this letter, we readdress this issue using a carefully compiled set of $B-R$ colors and $H_R(\alpha)$ magnitudes 
(as proxy for size) for 253 objects, including data for 10 new small objects.

We find that the bimodal behavior seen among Centaurs is a size related phenomenon, common to both Centaurs and small KBOs, 
\ie independent of dynamical classification.   
Further, we find that large KBOs also exhibit a bimodal behavior of surface colors, albeit distinct from the small objects and 
strongly dependent on the `Haumea collisional family' objects.
When plotted in $B-R$, $H_R(\alpha)$ space, the colors of Centaurs and KBOs display a peculiar $\mathcal{N}$ shape. 
\\
\\
{\bf Key words.} Kuiper belt: general

\vspace*{2ex}
\end{minipage}
\end{center}
}] 

%%%%%%%%%%%%%%%%%%%%%%%%%%%%%%%%%%%%%%%%%%%%%%%%%%%%%%%%%%%%%%%%%%%%%%%%%%%%
%%%%%%%%%%%%%%%%%%%%%%%%%%%%%%%%%%%%%%%%%%%%%%%%%%%%%%%%%%%%%%%%%%%%%%%%%%%%

\section{Introduction}

Discovered just 20 years ago \citep{1993Natur.362..730J}, the Kuiper Belt holds a vast
population of icy bodies orbiting the Sun beyond Neptune. Stored at very low
temperatures ($\sim$30-50~K), the Kuiper Belt Objects (KBOs) are expected to be well-preserved fossil
remnants of the solar system formation. Presently, $\sim$1600 KBOs have been identified and classified into several dynamical
families \citep[see Appendix \ref{app:database} and][for a review]{2008ssbn.book...43G}. KBOs which dynamically
evolve to become Jupiter Family Comets (JFCs) form a transient population, the
Centaurs, with short-lived chaotic orbits between Jupiter and Neptune
\citep{Kow77,1980MNRAS.192..481F,1997Icar..127...13L}. 

Between 1998 and 2003, we witnessed a debate on the surface colors of
KBOs and Centaurs. One team used very accurate surface colors and detected that
KBOs were separated into two distinct color groups
\citep{1998Natur.392...49T,2000Natur.407..979T,2003Icar..161..181T}. Other teams did not find evidence for such
color bimodality \citep{1999Icar..142..476B,2001AJ....122.2099J, 2002A&A...389..641H}. Careful reanalysis of the
data by \cite{2003A&A...410L..29P} indicated that only the Centaurs display bimodal colors,
\ie they are distributed in two distinct color groups, one with neutral
solar-like colors, and one with very red colors. KBOs on the other hand exhibit
a broad continuous color distribution, from neutral to very red, with no
statistical evidence for a color gap between the extrema \citep[][for a
review]{2008ssbn.book..105T}.

The relevance of this controversy lays on two possible interpretations: i) KBOs
and Centaurs are composed of intrinsically different objects, with distinct
compositions, which probably formed at different locations of the protosolar
disk, ii) KBOs and Centaurs are originally similar but evolutionary processes
altered them differently, hence their color diversity.  Most research focused
on the latter hypothesis, offering little improvement on our understanding of
the color distributions. 
\cite{1996AJ....112.2310L} proposed that the competition between a reddening effect of
irradiation of surface ices \citep{1987JGR....9214933T} and a bluing effect due
to collisional induced resurfacing of fresh, non-irradiated, ices might generate
the observed surface colors. The same authors, however, rejected this model as
being the primary cause of the color diversity, due to the lack of predicted
rotational color variations \citep{2001AJ....122.2099J}. Based on the same processes,
\cite{2002P&SS...50...57G}  proposed a more complex treatment of the irradiation process, by
implying an intricate structure of differently irradiated subsurface layers.
However, the collisional resurfacing effects became very hard to model, thus
making it very hard to provide testable predictions.  Later, \cite{2003Icar..162...27T}
showed that the collisional energies involved in different parts of the Kuiper
Belt did not corroborate the possible link between surface colors and
non-disruptive collisions. 

\cite{2004A&A...417.1145D} refined the first-mentioned model by considering the effects of a
possible cometary activity triggered by collisions, and a
size/gravity-dependent resurfacing. Cometary activity can modify the surface
properties through the creation of a neutral-color dust mantle.  \cite{2002AJ....123.1039J}
suggested that this process could explain why no JFCs are found with the
ultra-red surfaces seen in about half of the Centaurs. It has also been proposed that the
sublimation loss of surface ice from a mixture with red materials may be
sufficient to make the red material undetectable in the visible wavelengths
\citep{2009Icar..199..560G}. These might explain the Centaur color bimodality, 
as long as all were red when migrating inwards from the Kuiper Belt.
Although promising, these models did not provide an explanation for the color
bimodality of Centaurs, as they fail to reproduce the bluest colors observed
and their frequency. 

%%%%%%%%%%%%%%%%%%%%%%%%%%%%%%%%%%%%%%%%%%%%%%%%%%%
%%%%%%%%%%%%%%%%%%%%%%%%%%%%%%%%%%%%%%%%%%%%%%%%%%%

\section{Motivation for This Work}

We find it striking that the objects with both perihelion and semi-major axis between Jupiter and 
Neptune's orbits, the Centaurs --- by definition---, would display a different color distribution than 
physically and chemically similar objects with a semi-major axis 
slightly beyond Neptune's orbit, as is the case for Scattered Disk Objects (SDOs), for instance, or any other KBOs.
There is no evident physical consideration that would explain
the apparently sudden `transition' in surface color behavior (from bimodal to unimodal) precisely at 
Neptune's orbital semi-major axis $a_N$ =30.07~AU.
This difference between Centaurs and KBOs is particularly puzzling
since there is neither a sharp dynamical separation between them,
(the definition is somewhat arbitrary), nor a 
clearly identified family of KBOs in their origin. Although SDOs are frequently 
considered as the main source of Centaurs, Neptune Trojans, Plutinos, and Classical KBOs have been 
demonstrated as viable contributors \citep[][respectivelly]{2010MNRAS.402...13H, YuTre99, 2008ApJ...687..714V}. 
Further, Centaurs possess short dynamical lifetimes of $\sim5\cdot10^5-3\cdot10^7$ yr before being injected as JFCs or 
ejected again to the outer Solar System \citep{2004MNRAS.354..798H}. 
If some surface evolution mechanism, dependent on heliocentric distance, is responsible for the bimodal behavior of Centaurs, 
it must be acting extremely fast such that no intermediate colors are ever seen among them. 
Besides surface color bimodality, the most distinctive characteristic of Centaurs compared to `other' KBOs is their small size. 
Known KBOs are mostly larger than Centaurs,  simply because they are more distant and thus smaller objects are harder to detect. 

In this work, we address the
issue of the color distributions of Centaurs and KBOs. We present new data on
seven intrinsically faint (thus small) KBOs and three Centaurs, combined with a
new compilation of 253 published $B-R$ colors, and available 
$m_R(1,1,\alpha)$ magnitudes, or $H_R(\alpha)$, \ie absolute magnitude non-corrected from phase effects, and 
some identified spectral features.  
We study this large
sample of colors (including objects from all dynamical families) versus
absolute magnitude as a proxy for size,  with the implicit assumption that 
surface colors are independent of dynamical classification.
We present the most relevant results,
found in $B-R$ \vs $H_R(\alpha)$ space.

%%%%%%%%%%%%%%%%%%%%%%%%%%%%%%%%%%%%%%%%%%%%%%%%%%%
%%%%%%%%%%%%%%%%%%%%%%%%%%%%%%%%%%%%%%%%%%%%%%%%%%%

\section{Observations and Data Reduction}
\label{sec:obs}

Observations of 7 
KBOs and 1 Centaur were taken at the 8.2 m Subaru telescope, on 2008--07--02,
using $0.''206/$pix FOCAS camera in imaging mode with $2\times2$ binning
\citep[2 CCDs of $2048\times4096$ pixels,][]{2002PASJ...54..819K}.  Weather was
clear with seeing $\sim$$0.7''$.  We used the University of Hawaii UH 2.2 m
telescope, to observe 2 Centaurs on 2008--09--29, with the $0.''22/$pixel
Tektronix $2048\times 2048$ pixels CCD camera.  Weather was clear with seeing
$\sim$$0.9''$.  Both telescopes are on Mauna Kea, Hawaii, USA.  Images from
both instruments were processed using IRAF's CCDRED package following the
standard techniques of median bias subtraction and median flat-fielding
normalization.  

%%%%%%%%%%%%%%%%%%%%%%%%%%%%%%%%%%%%%%%%%%%%%%%%%%%
\begin{table}
\caption{Filters specifications}
\label{tab:filters}
\begin{center}
\begin{tabular}{ccccccc}
\hline
\hline
             & & \multicolumn{2}{c}{\bf 8.2m Subaru} & &  \multicolumn{2}{c}{\bf UH 2.2m} \\
\cline{3-4}  \cline{6-7}
Filter	  & & \multicolumn{2}{r}{\rule[-1mm]{0mm}{5mm} Wavelength (\AA)} & & \multicolumn{2}{r}{Wavelength (\AA)} \\
            & & Center & Width   & & Center & Width   \\
B         & & 4400    & 1080    & & 4480    & 1077 \\
R         & & 6600    & 1170    & & 6460    & 1245	 \\
\hline
\end{tabular}
\end{center}
\end{table}
%%%%%%%%%%%%%%%%%%%%%%%%%%%%%%%%%%%%%%%%%%%%%%%%%%%

Standard calibration was made observing 
Landolt standard stars \citep{1992AJ....104..340L} at different airmasses for each filter, 
obtaining the corresponding zeropoints, solving by non-linear least-square fits the transformation 
equations, directly in order of $R$ and $(B-R)$, using IRAF's PHOTCAL package. 
The characteristics of 
the filters used on each telescope were essentially equal (Tab. \ref{tab:filters}).
Subaru's data was calibrated using Landolt standard stars: 
107-612, PG1047+003B, 110-230, Mark A2, and 113-337, taken repeatedly at different airmasses. 
UH2.2m's data was calibrated, analogously, using the stars: 92-410, 92-412, 94-401, 94-394, 
PG2213-006A and PG2213-006B. These stars have high photometric accuracy and colors 
close to those of the Sun. 
We have used the typical extinction values
for Mauna Kea, $k_B=0.19$, and $k_R=0.09$ \citep[][ and CFHT Info Bulletin \#19]{1987PASP...99..887K}.
All fits had residuals $rms<0.02$, which were added quadratically to the photometric error on each 
measurement. Targets were observed twice in B and twice in R bands, to avoid object trailing in one long exposure. 
Each two B or R exposures were 
co-added centered in the object, and also co-added centered on the background stars. The former were used to measure 
the object, the latter to compute the growth-curve correction. The time and airmass of observation were computed 
to the center of the total exposure. 
We applied growth-curve correction techniques to
measure the target's magnitudes using IRAF's MKAPFILE task \cite[for details,
see][]{2004Icar..170..153P}.  Observation circumstances and results are 
shown in Table \ref{tab:obs}. 

%%%%%%%%%%%%%%%%%%%%%%%%%%%%%%%%%%%%%%%%%%%%%%%%%%%
\begin{table*}
\tabcolsep 4.5pt
\caption{Observational circumstances and photometric results of this work's data}
\label{tab:obs}
\begin{center}
\begin{tabular}{rlccccccccc}
\hline
\hline
\multicolumn{2}{c}{Object} & Dyn. Class$^*$ & Telescope & UT Date & $r$[AU] & $\Delta$[AU] & $\alpha[^{\circ}]$ & R & B-R & $H_R(\alpha)$ \\
\hline
(130391) & 2000~JG$_{81}$ & 2:1  & Subaru & 20080702UT07:24:58 & 34.073 & 34.817 & 1.2 & 23.12$\pm$0.03 & 1.42$\pm$0.06 & 7.75$\pm$0.06\\
(136120) & 2003~LG$_{7}$ & 3:1  & Subaru & 20080702UT09:42:53 & 32.815 & 33.659 & 1.0 & 23.54$\pm$0.05 & 1.27$\pm$0.09 &  8.32$\pm$0.05\\
(149560) & 2003~QZ$_{91}$ & SDO  & Subaru & 20080702UT13:08:33 & 25.849 & 26.509 & 1.7 & 22.48$\pm$0.03 & 1.30$\pm$0.05 & 8.30$\pm$0.03\\
                  & 2006~RJ$_{103}$ & Nep. Trojan  & Subaru & 20080702UT14:07:50 & 30.760 & 30.534 & 1.9 & 22.27$\pm$0.02 & 1.90$\pm$0.04 & 7.40$\pm$0.02\\
                  & 2006~SQ$_{372}$ & SDO  & Subaru & 20080702UT11:45:34 & 23.650 & 24.287 & 1.9 & 21.55$\pm$0.02 & 1.78$\pm$0.03 & 7.71$\pm$0.05 \\
                  & 2007~JK$_{43}$& SDO  & Subaru & 20080702UT08:08:13 & 23.113 & 23.766 & 1.9 & 20.73$\pm$0.02 & 1.40$\pm$0.03 & 7.03$\pm$0.02\\
                  & 2007~NC$_{7}$ & SDO  & Subaru & 20080702UT11:30:49 & 20.090 & 20.916 & 1.7 & 21.19$\pm$0.02 & 1.28$\pm$0.03 & 8.07$\pm$0.02 \\
(281371) & 2008~FC$_{76} $ & Cent & Subaru & 20080702UT11:13:05 & 11.119 & 11.793 & 3.8 & 19.79 $\pm$0.02 & 1.76$\pm$0.02 & 9.18$\pm$0.04\\
                  & 2007~RH$_{283}$ & Cent & UH2.2m & 20080929UT12:43:47 & 17.081 & 17.956 & 1.6 & 20.85 $\pm$0.03 & 1.20$\pm$0.05 & \\
                  & 2007~RH$_{283}$ & Cent & UH2.2m & 20080929UT12:57:51 & 17.081 & 17.956 & 1.6 & 20.90$\pm$0.03 & 1.28$\pm$0.06 & \\
                  &     mean...             &           &                &                                           &              &               &        &                                  & 1.24$\pm$0.07   & 8.44$\pm$ 0.04 \\
                  & 2007~UM$_{126}$ & Cent & UH2.2m & 20080929UT08:56:52 & 10.191 & 11.177 & 0.9 & 20.43$\pm$0.03 & 1.21$\pm$0.05 & \\
                  & 2007~UM$_{126}$ & Cent & UH2.2m & 20080929UT09:06:41 & 10.191 & 11.177 & 0.9 & 20.53$\pm$0.03 & 0.92$\pm$0.04 & \\
                  & 2007~UM$_{126}$ & Cent & UH2.2m & 20080929UT09:16:17 & 10.191 & 11.177 & 0.9 & 20.38$\pm$0.02 & 1.12$\pm$0.04 & \\
                  &     mean...             &           &                &                                           &              &               &        &                                 & 1.08$\pm$0.10   & 10.16$\pm$0.04 \\
\hline
\end{tabular}
\tablefoot{$^*$ Dynamical classes are: Centaur, Scattered Disk Object (SDO), Neptune Trojan (object located in 1:1 mean motion resonance with Neptune), 2:1, and 3:1, (objects located in 2:1 or 3:1 mean motion resonance with Neptune, respectively). For details on the classification see Appendix \ref{app:database}}
\end{center}
\label{default}
\end{table*}
%%%%%%%%%%%%%%%%%%%%%%%%%%%%%%%%%%%%%%%%%%%%%%%%%%%
%%%%%%%%%%%%%%%%%%%%%%%%%%%%%%%%%%%%%%%%%%%%%%%%%%%

\section{Compilation of Data}
\label{sec:compilation}

We compiled the visible colors for 290 objects (KBOs, Centaurs, and
Neptune Trojans) for which the absolute magnitude in R or V band was
accessible (\eg with individual magnitudes {\it and} observing date
available), and surface spectra information for 48 objects, as published in the literature to date (Feb. 2012). 
We computed the absolute magnitude 
$H_R(\alpha)\equiv m_R(1,1,\alpha) = R - 5\,log\,(r\cdot\Delta)$, 
where $R$ is the R-band magnitude, $r$ and $\Delta$ are
the helio- and geocentric distances in AU, respectively. In this
compilation, 253 objects have $B-R$ color available which is the
focus of this paper (see Table \ref{onlinetab:colors}), and 48 have also spectral information. The
description of the compilation method is presented in Appendix \ref{app:database}.
Sun-Object-Earth phase angles $\alpha$ are, typically, less than $1.5^{\circ}$ for KBOs and less than 
$4^{\circ}$ for Centaurs. Measurements of magnitude dependences on the phase angle for these objects, \ie phase coefficients $\beta$[mag/$\,^{\circ}$], 
are scarce but, so far, do not show evidence for extreme variability presenting an average value of $\beta=0.11\pm0.05$ \citep{2008ssbn.book..115B}.  
From the linear approximation $H_R(\alpha=0^{\circ})\approx H_R(\alpha)-\alpha \beta$, by not correcting the absolute magnitude from phase effects 
we are slightly overestimating it. We will deal with this issue in Sec. \ref{sec:doublebimodality}. 

Recent works have shown that there is no strong correlation between object diameter $D$ and geometric albedo $p_V$, nor between geometric albedo $p_V$ and absolute magnitude $H_R$ \citep{2008ssbn.book..161S, 2012arXiv1202.1481S, Vilenius+2012, Mommert+2012}. However, from the 74 diameter and albedo measurements of Centaurs and KBOs made using Herschel and/or Spitzer telescopes, published in the aforementioned works, 
we verify that $H_R$ and $D$ correlate very strongly with a Spearman-rank correlation of $\rho=-0.92^{+0.03}_{-0.02}$, with a significance level $SL\ll 0.01\%$ \citep[error bars computed using bootstraps, for details see][]{2007AJ....134.2186D}. Consequently, absolute magnitude
is a very good proxy for size. 

%%%%%%%%%%%%%%%%%%%%%%%%%%%%%%%%%%%%%%%%%%%%%%%%%%%

\section{An $\mathcal{N}$-shaped Doubly Bimodal Structure}
\label{sec:doublebimodality}

%%%%%%%%%%%%%%%%%%%%%%%%%%%%%%%%%%%%%%%%%%%%%%%%%%%
\begin{figure} 
\resizebox{\hsize}{!}{\includegraphics{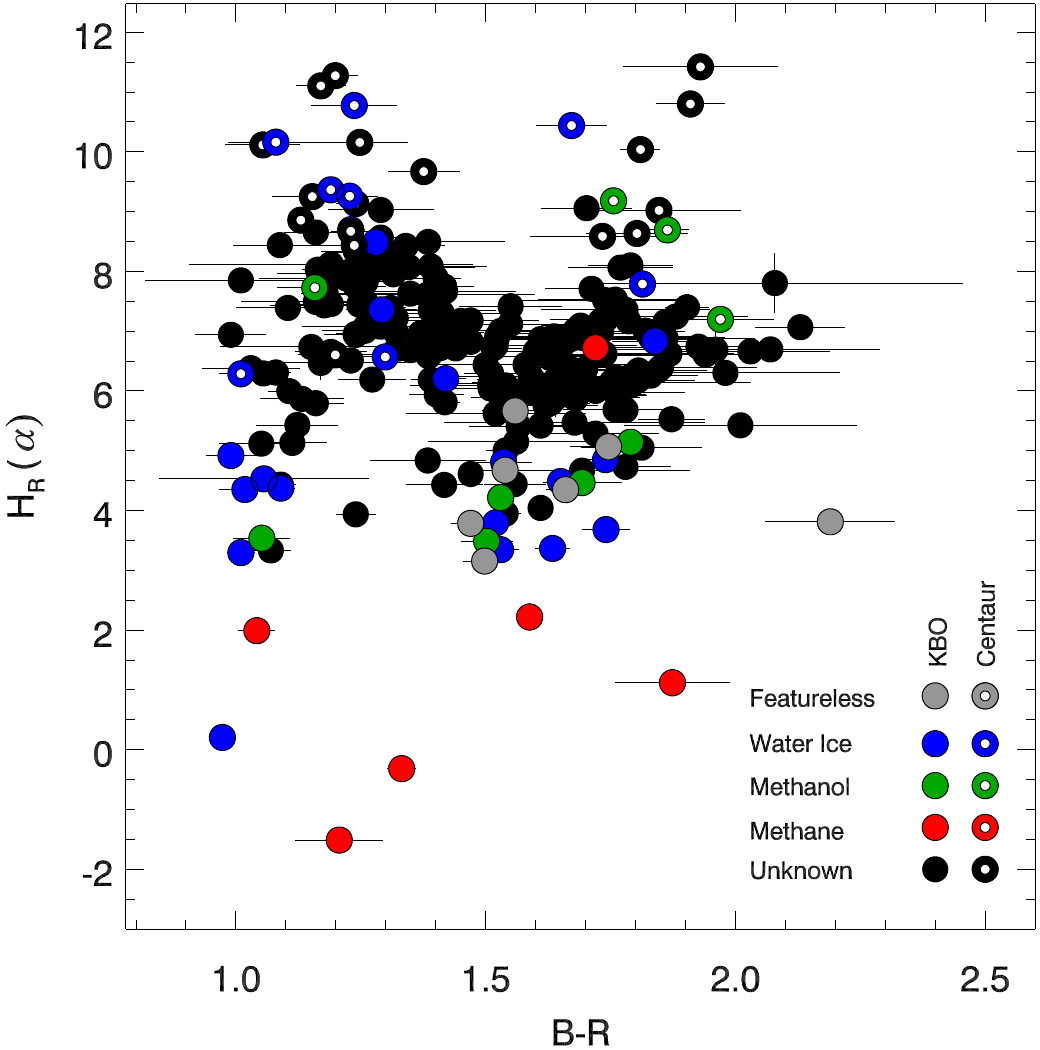}} 
\caption{$B-R$ \vs $H_R(\alpha)$ plot of all 253 objects. KBOs are represented by solid circles 
and Centaurs by white dotted solid circles. 
Objects with $H_R(\alpha)\geqslant6.8$ separate into two color groups with a `gap' centered at $B-R\sim1.60$. 
Objects with $H_R(\alpha)\leqslant5.0$ also show statistical evidence for separation in two colors groups 
but with a `gap' centered at $B-R\sim1.25$.
Objects spectra with known features of water ice, methane, methanol, and featureless spectra, 
are coded using colors as described in the legend. 
There is no obvious/clear connection between $B-R$ colors and 
the presence of spectral features. 
}
\label{fig:HBR}
\end{figure}
%%%%%%%%%%%%%%%%%%%%%%%%%%%%%%%%%%%%%%%%%%%%%%%%%%%

In Fig. \ref{fig:HBR} we plot $R$-band absolute magnitude $H_R(\alpha)$ (proxy for
object's size) against $B-R$ color for all ($n=253$) objects in our database. The
cloud of points forms a recognizable $\mathcal{N}$ shape with an apparent double bimodal structure in
color. The smaller objects (upper part of the plot) show a bimodal $B-R$
distribution.  Although apparently dominated by Centaurs, this bimodal distribution also
includes KBOs of similar $H_R(\alpha)$, which suggests that {\bf the bimodal structure in
$B-R$ color is a property of the smaller objects in general, regardless of
their dynamical family}. This bimodality appears do disappear for objects with $H_R(\alpha)\lesssim7$ 
where the $B-R$ color distribution seems unimodal.
Most interestingly, we note that towards the larger objects (lower part of the
plot) the colors suggest the presence of another bimodal behavior, with the gap between the two groups
shifted towards the blue with respect to the `small' object bimodality.  This
new `large' object bimodality is explicitly reported for the first time.

When performing hypotheses testing one should adopt a critical value of significance $\alpha$.
The value $\alpha$ is the maximum probability (risk) we are willing to take in rejecting the null hypothesis $H_0$ 
(\ie to claim no evidence for bimodality) when it is actually true (\ie data is truly bimodal/multimodal) 
--- also called type I error probability. Such value is often a source of debate, 
as the theories of hypotheses testing themselves \citep[\eg][]{Lehmann1993}.
The decision relies mostly whether the effects of a right or wrong decision are of practical importance 
or consequence. The paradigm is: by diminishing the probability of wrongly reject a null hypothesis 
(\eg decide for bimodality when bimodality was not present in the parent population) we increase the 
probability of wrongly accepting the null hypothesis (\ie deciding for unimodality when 
bimodality was in fact present), also called type II error probability, or risk factor $\beta$. 
Some authors and/or research fields, consider that there is only sufficient evidence against $H_0$ when the 
achieved significance level is $SL< 0.3\%$, \ie using $\alpha=0.3\%$ (the $3\sigma$ Gaussian probability), 
others require even $\alpha=0.0003\%$ ($6\sigma$). 
Such might be a criterion for rejection of $H_0$ but not a very useful `rating' for the evidence against $H_0$, 
which is what we are implicitly doing. 
We rate the evidence against $H_0$ following a most common procedure in Statistics: $SL< 5\%$ --- 
reasonably strong evidence against $H_0$,  $SL<2.5\%$ --- strong evidence against  $H_0$, 
and $SL<1\%$ --- very strong evidence against $H_0$ \citep[\eg][]{EfrTib93}, adding also the 
common procedure in Physics: $SL<0.3\%$ --- clear evidence against $H_0$. Further, for better readability, 
throughout this work we may employ the abuse of language `evidence for bimodality' 
instead of the statistically correct term `evidence against unimodality'. 

\begin{figure} 
\resizebox{\hsize}{!}{\includegraphics{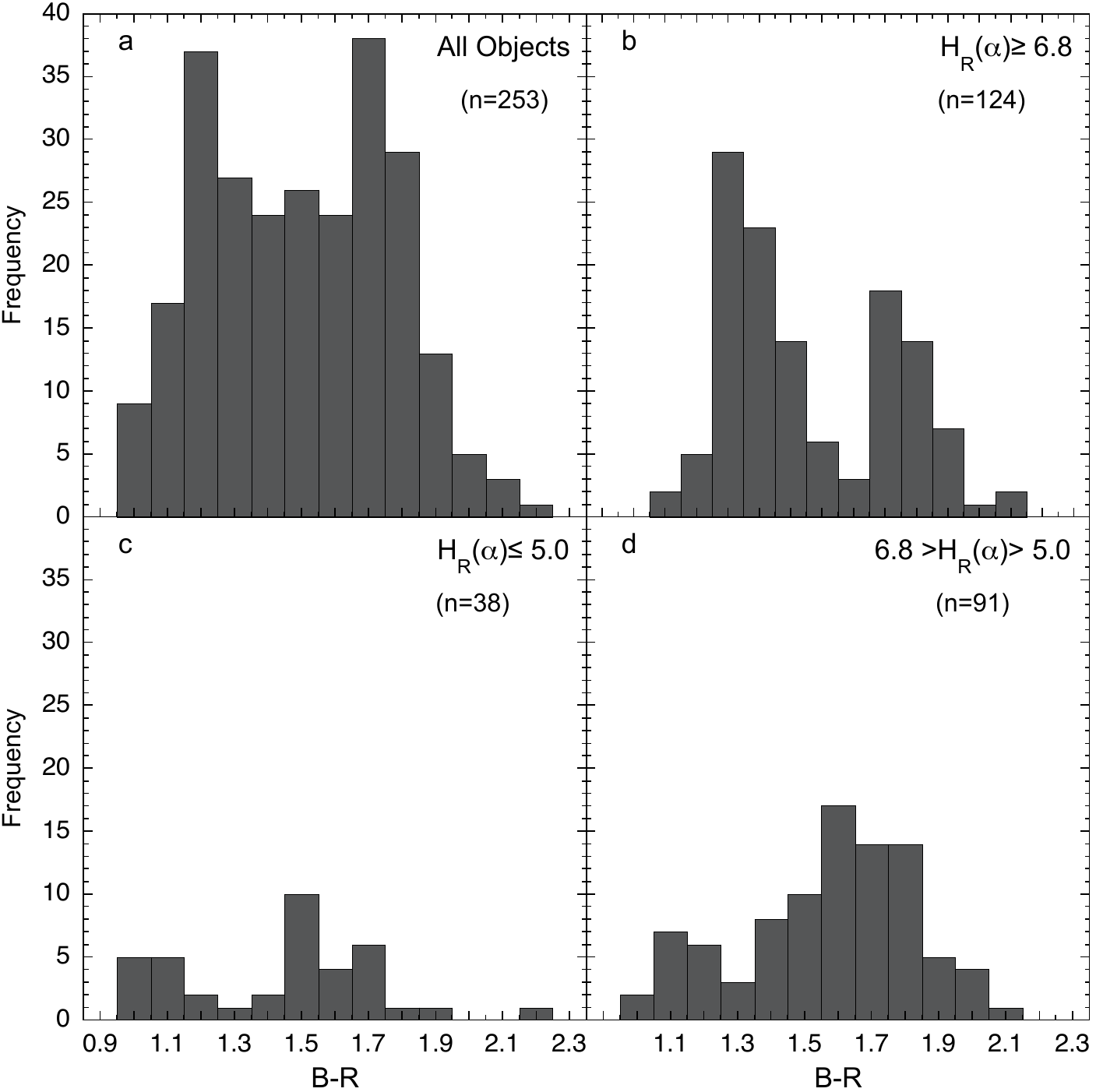}} 
\caption{Histograms of $B-R$ colors from selected $H_R(\alpha)$ ranges: 
a) All the 253 objects. Taken globally do not exhibit statistical evidence 
for bimodality, which was known to exist among Centaurs.
b) The 124 `small' objects, with $H_R(\alpha)\geqslant 6.8$. 
Evidence for bimodal behavior is clear and still present when removing Centaurs. 
c) The 38 `large' objects, with $H_R(\alpha)\leqslant 5.0$. A bimodal behavior is shown but
it loses the statistical significance without the `Haumea collisional family' objects.
d) The 91 `intermediate' size objects, $6.8>H_R(\alpha)>5.0$. Regardless of the apparent 
small gap at  $B-R\sim1.3$ there is no statistical 
evidence for two separate groups. 
}
\label{fig:histograms}
\end{figure}

Using the R software's \citep[version 2.14.1;][]{R-cran} Dip Test package
\citep{Hart85, HartHart85, R-diptest} we test the null hypothesis $H_0$: 
`the sample is consistent with an unimodal parent distribution' over all 
objects in the $B-R$ \vs $H_R(\alpha)$ space, against the alternative hypothesis $H_1$: 
`the sample is not consistent with an unimodal parent distribution' (hence it is bimodal or multimodal).
The full sample, in spite of the apparent two spikes, shows no relevant evidence against color unimodality,
neither with ($n=253$, $SL=17\%$) nor without ($n=224$, $SL=41\%$)
Centaurs (see Fig. \ref{fig:histograms} a).  
The Centaur population ($n=29$) shows strong evidence against unimodality at $1.6\%$. Removing the 3
brightest Centaurs (with $H_R(\alpha)\gtrsim6.6$) improves the significance to $0.3\%$.  To refine the
analysis and test different ranges in $H_R(\alpha)$  we ran the Dip Test on
sub-samples using a running cutoff in $H_R(\alpha)$ that was shifted by 0.1 mag between
consecutive tests. 

%%%%%%%%%%%%%%%%%%%%%%%%%%%%%%%%%%%%%%%%%%%%%%%%%%%
\paragraph{\bf Bimodal distribution of `small' objects:} We performed iterative
Dip Tests with a $H_{R:\,cut}$ starting at the maximum $H_R(\alpha)$ value, and
decreasing in steps of 0.1 mag; in each iteration we run the test on those
objects above the cutoff line (\ie with $H_R(\alpha)\geqslant H_{R:\,cut}$). We stop
shifting $H_{R:\,cut}$ when we detect the maximum of evidence against unimodality  
(\ie a minimum of significance level, henceforth accepting the alternate hypothesis `the distribution is bimodal/multimodal') 
Evidence for bimodality at significance levels better than $5\%$ start to be seen for objects with $H_R(\alpha)\geqslant7.1$. 
This evidence peaks at a significance of $0.1\%$ for the $124$
faint objects with $H_R(\alpha)\geqslant6.8$.  

We propose that the visible surface color distribution of (non-active) icy bodies of the outer Solar System depends only on objects size, and is independent of their dynamical classification. No mechanism has yet been found to explain the color bimodality only for Centaurs. However, since such mechanism might exist even if not yet found, 
we re-analyze the sample removing the Centaurs. Naturally, the sampling of the smaller objects diminishes considerably, 
hence reducing the statistical significance against the null hypothesis (\ie increases the probability of 
observing two groups on a purely random distribution of colors).
Nonetheless, the $98$ remaining objects with $H_R(\alpha)\geqslant6.8$ show evidence for bimodality at a significance level of $3.5\%$, 
reaching a significance minimum of $1.8\%$ for the $165$ objects with 
$H_R(\alpha)\geqslant5.8$. In both cases the `gap' is centered around $B-R\sim1.60$ 
(see Figs. \ref{fig:HBR} and \ref{fig:histograms} b).

%%%%%%%%%%%%%%%%%%%%%%%%%%%%%%%%%%%%%%%%%%%%%%%%%%%
\paragraph{\bf Bimodal distribution of `large' objects:} We test the brightest
part of the sample using a cutoff limit starting at the minimum $H_R(\alpha)$ value; we
consider objects below the cutoff (\ie brighter than $H_{R:\,cut}$) and shift
it up in steps of 0.1 mag. 
We find very strong evidence against unimodality for objects 
with $H_R(\alpha)\lesssim 5.0$ ($SL=0.9\%$). Data still shows reasonably strong evidence against 
unimodality for objects up to $H_R(\alpha)\lesssim5.6$. 
The `gap' is located at
$B-R\sim1.25$. There are no Centaurs in this brightness range. 
Explicitly,   
evidence for `large' objects bimodality has not been previously reported.
(see Figs. \ref{fig:HBR} and \ref{fig:histograms} c). Removing from the sample 
the 7 objects belonging to the `Haumea collisional family' \citep{2007Natur.446..294B,2010A&A...511A..72S}, 
all clustered on the lower left `leg' of the $\mathcal{N}$ shape, erases the statistical evidence against the 
null hypothesis, even if still suggestive to the eye.
Therefore, with the present data sample, the `evidence for bimodality' among bright KBOs 
cannot be stated as independent from the peculiar properties of the Haumea collisional family. 

%%%%%%%%%%%%%%%%%%%%%%%%%%%%%%%%%%%%%%%%%%%%%%%%%%%
\paragraph{\bf The `intermediate' size continuum:} The 91 objects with
$6.8>H_R(\alpha)>5.0$, which include 3 Centaurs, do not show evidence against a unimodal behavior 
($SL=98.0\%$) even if a small gap seems suggestive to the eye (see Figs. \ref{fig:HBR} and \ref{fig:histograms} d).
However, statistically, their inclusion in the fainter
group does not decrease the significance below the `strong evidence against unimodality', \ie 
$SL=2.5\%$ (see Figs. \ref{fig:HBR} and \ref{fig:histograms} d). On the other hand, if added to the `large'  
objects the statistical evidence for bimodality of `large' objects does not hold. 

%%%%%%%%%%%%%%%%%%%%%%%%%%%%%%%%%%%%%%%%%%%%%%%%%%%
%%%%%%%%%%%%%%%%%%%%%%%%%%%%%%%%%%%%%%%%%%%%%%%%%%%

\paragraph{}
To check for the effects of non-correcting $H_R(\alpha)$ from phase angle effects we performed 
Monte-Carlo simulations. First, we compute all the possible $\alpha$ values and their probability 
distribution for an `average' Centaur with semi-major axis $a=15$ AU. 
The maximum $\alpha$ is $3.8^{\circ}$ being the median value $3.2^{\circ}$.
Analogously, we do the same for a KBO with $a=40$ AU. The maximum $\alpha$ is $1.4^{\circ}$ 
and the median value $1.2^{\circ}$. Therefore, on average, our absolute magnitudes might be 
{\it overestimated} by $\Delta H_R \approx 0.35$, for Centaurs, and by  
$\Delta H_R \approx 0.13$, for KBOs. 
Simulating 1000 `phase-corrected' $H_R$ data-samples, following the probability 
distribution of the corresponding $\alpha$ angles did not alter any of the results obtained 
using simply $H_R(\alpha)$.

%%%%%%%%%%%%%%%%%%%%%%%%%%%%%%%%%%%%%%%%%%%%%%%%%%%
%%%%%%%%%%%%%%%%%%%%%%%%%%%%%%%%%%%%%%%%%%%%%%%%%%%

\section{Interpretation}

Our analysis shows that the $B-R$ colors of Centaurs and KBOs when plotted as a
function of $H_R(\alpha)$ display an N-shaped, double bimodal behavior. The color
distribution seems to depend on object size (intrinsic brightness) instead of 
dynamical family. Using the brightness-size-albedo relation
$D_{km}=2\sqrt{\,2.24 \cdot 10^{16} \cdot 10^{\,0.4\,(H_{R\,\odot}-H_R)}/p_R}$, 
with solar $H_{R\,\odot}=-27.10$, the main issue is to choose a canonical
geometric albedo value $p_R$. Recent works 
\citep{2008ssbn.book..161S, 2012arXiv1202.1481S, Vilenius+2012, Mommert+2012}
show a wide range of values, for each dynamical family, in some cases far from the 0.04 value previously 
assumed from comet studies. As we need only a rough estimate of size ranges, we pick the 
average value of $p_R=0.09$. Using this
parameter, objects with diameters $165\lesssim D_{km} \lesssim 380$ present a
rather continuous range of B-R colors.

Visible and near-infrared (NIR) spectroscopy for about 75 bright objects
\citep[][for a review]{2011Icar..214..297B} also indicates that the surface
compositions of KBOs and Centaurs is very diverse. The largest objects are
coated in methane ice, while intermediate size objects display water-ice
features, sometimes with traces of other volatiles.  Small KBOs generally
have featureless spectra.  The presence of volatiles on the surface of an
object may be related to its ability to retain them, \ie to its size and
temperature \citep{2007ApJ...659L..61S}. It should also depend on the
subsequent irradiation history \citep{2011ApJ...739L..60B}. However no
correlation can be made to date between visible colors and NIR spectral
properties. For example, two objects of comparable size, Quaoar and Orcus,
both exhibit water ice-dominated surfaces but have, respectively, very red and
neutral visible colors \citep{2010A&A...520A..40D}.

Objects smaller than $\sim$100-150~km, including most of the known Centaurs, are 
believed to be fragments from the collision of larger objects \citep{2005Icar..173..342P}. 
Predicting the properties of these fragments is a complex task, but the field shows 
promising advances \citep[for a review, see][]{2008ssbn.book..195L}. An immediate 
hypothesis is that red and neutral objects are the only possible outcomes of a disruptive 
collision. Thermal evolution modeling suggests that KBOs, especially large ones, should 
have a layered structure, including some liquid water leading to a complete differentiation 
of the object \citep{2006Icar..183..283M, 2011A&A...529A..71G}. A catastrophic collision 
could result in the formation of fragments with very different properties, depending on 
whether they come from the core of the parent body, or its mantle, or some subsurface 
layers. However, our current knowledge of KBOs internal properties and evolution is still 
incipient to support or discard such an hypothesis.  
Besides, it is hard to understand why objects with $B-R\sim1.6$ (in the gap of the small 
object's bimodal distribution) should be inexistent. 
Maybe their relative number is so small compared to the neutral and red groups that we 
can hardly observe them, leading to another puzzling question. 
Research on these aspects should be encouraged, in particular 
the detection and measurement of many more small objects --- KBOs and Centaurs --- could 
help further constraining their color distribution and other properties. 
The objects in the `intermediate' $H_R(\alpha)$ range ($6.8>H_R(\alpha)>5.0$) 
seem unimodally distributed in $B-R$ color; 
they might represent a transition phase between the two bimodal distributions.
These medium-sized objects are probably too large to be remnants from disruptive collisions, 
and too small to have recently undergone cryovolcanic activity (they may not even be differentiated).
They might, actually, represent the only group where the outcomes of the combined effects of 
different birthplaces, space weathering and thermal processing can be studied or analyzed.

The evidence for bimodal distribution among the largest objects is also puzzling. These are supposedly
the best studied objects, yet the evidence for a bimodal distribution of their surface colors has never been reported. 
Nonetheless, when removing the 7 Haumea collisional family objects from our sample it no longer provides 
evidence against an unimodal distribution, even if apparent to the eye. This issue should be further analyzed 
when larger sampling is available. 

In this work, we confirm that there is no 
noticeable link between the surface composition of an object and its visible colors. 
Objects hosting water ice are distributed both among large and small objects, 
and among red and blue ones. When it comes to volatiles such as methane (CH$_4$)
or methanol (CH$_3$OH), we find that they are also distributed among all groups, although 
they might be more difficult to detect on small/fainter objects. We nonetheless find a cluster 
of featureless objects among the red group of large objects: these might represent the most 
irradiated/oldest surfaces in the overall population. Therefore, it seems that a simple explanation 
such as the model of atmospheric escape proposed by \cite{2007ApJ...659L..61S} might not 
be sufficient to explain the colors and compositions of KBOs. The reason why they evolved 
in two different color groups can be very complex, and should involve different thermal, 
collisional, irradiation histories, on top of possible different birthplaces.

%%%%%%%%%%%%%%%%%%%%%%%%%%%%%%%%%%%%%%%%%%%%%%%%%%%
%%%%%%%%%%%%%%%%%%%%%%%%%%%%%%%%%%%%%%%%%%%%%%%%%%%

\section{Summary}

In this work we analyze the $B-R$ color distribution as a function of $H_R(\alpha)$ magnitude for 253 Centaurs 
and KBOs, including 10 new measurements, and with the information on their NIR spectral features. 
Using the known diameters, $D$, and albedos, $p_V$, of 74 of these objects 
we verify that $H_R$ and $D$ correlate very strongly ($\rho=-0.92^{+0.03}_{-0.02}$, $SL\ll 0.01\%$) 
validating $H_R$ as a good proxy for size. Further, through simulations, we show that not correcting $H_R(\alpha)$ 
to $H_R(\alpha=0^{\circ})$ does not change any of the global results. Our analysis shows:

\begin{enumerate}

\item The $B-R$ \vs $H_R(\alpha)$ color distribution is $\mathcal{N}$-shaped, evidencing that 
$B-R$ colors are probably dominated by a size effect independent from dynamical classification. 

\item Small objects, including both KBOs and Centaurs, display a bimodal structure of 
$B-R$ colors at $0.1\%$ significance level (\ie objects with $H_R(\alpha)\geqslant 6.8$, or $D_{km}\lesssim 165$, assuming $p_R=0.09$)
with the `gap' centered at $B-R\sim1.60$. Removing Centaurs from the sample reduces greatly the sampling on small 
objects reducing also the significance of the result to $3.8\%$.

\item Large objects evidence also for a bimodal structure, with minimum 
significance of $0.9\%$, for $H_R(\alpha)\lesssim5.0$ ($D_{km}\gtrsim 380$, assuming $p_R=0.09$), and color `gap' centered at  
$B-R\sim1.25$. Reasonable evidence for this bimodality starts when considering only objects with $H_R(\alpha)\lesssim5.6$ ($D_{km}\gtrsim 290$) 
dropping below the critical $5\%$ when reaching $H_R(\alpha)\lesssim4.4$ ($D_{km}\gtrsim 500$). However, this behavior seems  
dominated by the presence of 7 Haumea collisional family objects which `cluster' at the lower left leg of the $\mathcal{N}$-shape. 
Once removed, there is no statistical evidence against compatibility with a random unimodal distribution for the larger KBOs.

\item Intermediate sized objects do not show incompatibility with a continuum of $B-R$ colors (\ie 
$6.8>H_R(\alpha)>5.0$, or $165\lesssim D_{km} \lesssim 380$, assuming $p_R=0.09$). These objects 
seem too large to be remnants from disruptive collisions and too small to hold cryovolcanic activity. 
They might be the best targets to study the combined effects of different birthplaces, 
different space weathering, and different thermal processing. Further studies are encouraged.

\item Inspecting the NIR spectral properties against $B-R$ colors shows no obvious link between the  
colors and the chemical composition of the objects' surfaces.

\end{enumerate}

%%%%%%%%%%%%%%%%%%%%%%%%%%%%%%%%%%%%%%%%%%%%%%%%%%%
%%%%%%%%%%%%%%%%%%%%%%%%%%%%%%%%%%%%%%%%%%%%%%%%%%%

\begin{acknowledgements}
The authors thank Rachel Stevenson, Megan Bagley, and Takashi Hatori for assisting with the 
observations at Subaru telescope. NP was partially supported  by NASA's Origins grant to D. Jewitt, 
by the European Social Fund, by the Portuguese Foundation for Science and Technology 
(FCT, ref.: BPD/ 18729/ 2004), and by the transnational cooperation agreement 
FCT-Portugal / CNRS-France (ref.: 441.00). AGL was supported by a NASA Herschel grant to D. Jewitt. 
PL is grateful for financial support from a Michael West
Fellowship and from the Royal Society in the form of a Newton
Fellowship
\end{acknowledgements}

%%%%%%%%%%%%%%%%%%%%%%%%%%%%%%%%%%%%%%%%%%%%%%%%%%%
%%%%%%%%%%%%%%%%%%%%%%%%%%%%%%%%%%%%%%%%%%%%%%%%%%%

%% references
\bibliographystyle{aa}
%\bibliography{mnemonic,all_bib2012}

%%%%%%%%%%%%%%%%%%%%%%%%%%%%%%%%%%%%%%%%%%%%%%%%%%%
%%%%%%%%%%%%%%%%%%%%%%%%%%%%%%%%%%%%%%%%%%%%%%%%%%%
\begin{appendix} %Online appendix ... 
\section{Compiled Database}
\label{app:database}

For each object, we compute the average color index from the different
papers from data obtained {\it simultaneously} in B and R bands (\eg
contiguous observations within a same night). When individual R
apparent magnitude and date is available, we compute the 
$H_R(\alpha) = R - 5\,log\,(r\cdot\Delta)$, where $R$ is the R-band
magnitude, $r$ and $\Delta$ are the helio- and geocentric distances at
the time of observation in AU, respectively. When V and $V-R$ color is
available, we derive an R and then $H_R(\alpha)$ value. We do not correct for
the phase angle $\alpha$ effect as we need only a general estimation
of the absolute magnitude for our complete sample. In addition, few
objects have phase correction coefficient available in the literature,
and no universally accepted canonical values per dynamical class can
be strictly adopted. Table \ref{onlinetab:colors} presents the resulting
values. 
This table includes also spectral information on the presence of water ice, methanol, methane, or confirmed 
featureless spectra, as available in the literature. We highlight only the cases with clear bands on the spectrum which were 
reported/confirmed by some other work.

There is no strict definition for the dynamical classes of Centaurs and KBOs. 
Roughly speaking: objects orbiting in mean motion resonances with Neptune are called `resonants' 
(if located in the 1:1 resonance are also known as Neptune Trojans, and 
known as Plutinos if located in the 3:2 resonance); Centaurs are the objects 
with orbits between those of Jupiter and Neptune; Scattered Disk Objects (SDOs), are those within probable 
gravitational influence of Neptune; Detached KBOs, are those beyond past or future 
gravitational influence by Neptune; Classical KBOs , are those with rather circular orbits beyond 
Neptune and below the 2:1 resonance region (being called Hot if their orbital inclination is 
higher than $5^{\circ}$ or Cold if lower). 

To determine the dynamical class we first gathered the orbital elements, with epoch 2011--12--05, 
from `The Asteroid Orbital Elements Database', 
\verb+astorb.dat+ \footnote{ftp://ftp.lowell.edu/pub/elgb/astorb.dat.gz},  
maintained by the `Lowell Observatory'  
based on astrometric observations by the `Minor Planet Center'. 
Then, using the particular classification scheme suggested by 
\cite{2007Icar..189..213L}, including their analysis of objects located in the mean motion 
resonances (MMR) with Neptune, dynamical class was determined 
following a 11 steps algorithm:

\begin{enumerate}

\item $q<a_J \Rightarrow$ Not analysed
\item in $1:1$ MMR with Neptune $\Rightarrow$ Neptune Trojan
\item in $3:2$ MMR with Neptune $\Rightarrow$ Plutino
\item in other MMR with Neptune $\Rightarrow$ Other Resonant
\item $q>a_J \wedge a<a_N \Rightarrow$ Centaur
\item $a_J < q < a_N \wedge a\geqslant a_N \Rightarrow$ Scattered Disk Object (SDO)
\item $a_N < q \leqslant 37 \,AU  \Rightarrow$ Scattered Disk Object (SDO)
\item $q\geqslant 40 \,AU \wedge a\geqslant 48 \,AU  \Rightarrow$ Detached KBO (DKBO)
\item $37 \,AU \leqslant q \leqslant 40 \,AU  \Rightarrow$ Scattered or Detached KBO (SDKBO)
\item $i<5^{\circ} \wedge \{ \,[ q\geqslant 37 \,AU \wedge ( 37 \,AU \leqslant a \leqslant 40 \,AU) ] \vee [q\geqslant 38 \,AU \wedge (42 \,AU \leqslant a \leqslant 48 \,AU)]\,\} \Rightarrow$ Cold Classical KBO (cCKBO)
\item $i\geqslant5^{\circ} \wedge q\geqslant 37 \,AU \wedge  ( 37 \,AU \leqslant a \leqslant 48 \,AU ) \Rightarrow$ Hot Classical KBO (hCKBO)

\end{enumerate}

\noindent
being $q$ and $a$ the object's perihelion and semi-major axis, respectively. Jupiter semi-major axis is $a_J$, and Neptune's is $a_N$. Note that throughout the algorithm an object can be reclassified. 

We are aware that there are more complex classification schemes, which may be more refined, but the boundaries between families do 
not change significantly.  We chose this one for its computational simplicity. 

References for the colors presented in Table \ref{onlinetab:colors} are :
(1) \cite{1996AJ....112.2310L}; (2) \cite{1997P&SS...45.1607L}; (3)
\cite{2003A&A...400..369R}; (4) \cite{1997AJ....113.1893R}; (5)
\cite{2002Icar..160...59R}; (6) \cite{1998Natur.392...49T}; (7)
\cite{2001AJ....122.2099J}; (8) \cite{2002AJ....124.2279D}; (9)
\cite{2000Natur.407..979T}; (10) \cite{2001A&A...380..347D}; (11)
\cite{1997Icar..126..212T}; (12) \cite{1998AJ....115.1667J}; (13)
\cite{1999Icar..142..476B}; (14) \cite{2001A&A...378..653B}; (15)
\cite{2007AJ....134.2186D}; (16) \cite{2001Icar..154..277D}; (17)
\cite{1997MNRAS.290..186G}; (18) \cite{2002A&A...395..297B}; (19)
\cite{2003Icar..161..181T}; (20) \cite{2000A&A...356.1076H}; (21)
\cite{2010AJ....139.1394S}; (22) \cite{2000AJ....120..496B}; (23)
\cite{2008AJ....136.1502R}; (24) Tegler et
al. http)//www.physics.nau.edu/~tegler/research/survey.htm; (25)
\cite{2003ApJ...599L..49T}; (26) \cite{2001A&A...371..753P}; (27)
\cite{2002ApJ...566L.125T}; (28) \cite{2004Icar..170..153P}; (29)
\cite{2001ApJ...548L.243F}; (30) \cite{2005Icar..174...90D}; (31)
\cite{2009A&A...494..693S}; (32) \cite{2003Icar..162..408D}; (33)
\cite{2004A&A...421..353F}; (34) \cite{2005P&SS...53.1501D}; (35)
\cite{2006Icar..183..168G}; (36) \cite{2007AJ....133...26R}; (37)
\cite{2010AJ....140...29R}; (38) \cite{2006ApJ...639.1238R}; (39)
\cite{2008AJ....135.1749L}; (40) \cite{2010A&A...511A..72S}; (41)
\cite{2006Sci...313..511S}.

References for the spectral features indicated in Table \ref{onlinetab:colors} are:
(a) \cite{2003A&A...400..369R}; 
(b) \cite{1998Icar..135..389C}; 
(c) \cite{2000ApJ...542L.155K}; 
(d) \cite{2009A&A...501..777G}; 
(e) \cite{2001AJ....122.2099J}; 
(f) \cite{1999ApJ...519L.101B}; 
(g) \cite{2008AJ....135...55B}; 
(h) \cite{2009Icar..201..272G}; 
(i) \cite{2011Icar..214..297B}; 
(j) \cite{2010A&A...521A..35D}; 
(k) \cite{2005AJ....130.1299G}; 
(l) \cite{2010AJ....140.2095B}; 
(m) \cite{2010A&A...520A..40D}; 
(n) \cite{2009A&A...496..547P}; 
(o) \cite{2009AJ....137..315M}; 
(p) \cite{2007AJ....133..284B}; 
(q) \cite{2008ApJ...684L.107S}.

%%%%%%%%%%%%%%%%%%%%%%%%%%%%%%%%%%%%%%%%%%%%%%%%%%%

\begin{center}
\setlength{\tabcolsep}{1mm}
\begin{longtable}{llccll}
\caption{\label{onlinetab:colors}Compilation of absolute magnitude $H_R(\alpha)$, $B-R$ colors,  and spectral features used in this work}\\
\hline\hline
Object              & Dynamical Class  & $H_R(\alpha)$   & $B-R$  & Spectral features & References\\
\hline
\endfirsthead
Table \ref{onlinetab:colors} cont'd\\
\hline\hline
Object              & Dynamical Class  & $H_R(\alpha)$   & $B-R$  & Spectral features & References\\
\hline
\endhead
\hline
\multicolumn{6}{l}{\footnotesize {\bf References:} see Appendix \ref{app:database}} \\
\endfoot
(2060) Chiron            &     Centaur                          &  6.287$\pm$0.022   & 1.010$\pm$0.044  & Water ice &1, 2, 3, a \\                             
(5145) Pholus            &     Centaur                          &  7.198$\pm$0.056   & 1.970$\pm$0.108  & Methanol  &4, b   \\                                 
(7066) Nessus            &     Centaur                          &  9.020$\pm$0.068   & 1.847$\pm$0.165  & &1\\
(8405) Asbolus           &     Centaur                          &  9.257$\pm$0.120   & 1.228$\pm$0.057  & Water ice &4, 5, c  \\                               
(10199) Chariklo         &     Centaur                          &  6.569$\pm$0.015   & 1.299$\pm$0.065  & Water ice &6, 7, d  \\                              
(10370) Hylonome         &     Centaur                          &  9.250$\pm$0.131   & 1.153$\pm$0.081  & & 1, 6, 8\\
(15760) 1992~QB$_1$      &     Cold Classical                   &  6.867$\pm$0.121   & 1.670$\pm$0.145  & & 1, 7, 9 \\
(15788) 1993~SB          &     Plutino                          &  8.032$\pm$0.122   & 1.276$\pm$0.100  & & 7, 9, 10\\
(15789) 1993~SC          &     Plutino                          &  6.722$\pm$0.074   & 1.720$\pm$0.140  & Methane & 1, 7, 11, 12, e \\         
(15810) 1994~JR$_1$      &     Plutino                          &  6.867$\pm$0.077   & 1.610$\pm$0.216  & & 13\\
(15820) 1994~TB          &     Plutino                          &  7.527$\pm$0.091   & 1.759$\pm$0.155  & &1, 7, 10, 11, 13\\
(15874) 1996~TL$_{66}$   &     Scattered Disk Object            &  5.131$\pm$0.144   & 1.113$\pm$0.070  & &6, 7, 12, 13, 14\\ 
(15875) 1996~TP$_{66}$   &     Plutino                          &  6.953$\pm$0.071   & 1.678$\pm$0.123  & &6, 7, 12, 13, 14, 15\\ 
(15883) 1997~CR$_{29}$   &     Scattered Disk Object            &  7.076$\pm$0.135   & 1.260$\pm$0.128  & &7, 16\\ 
(16684) 1994~JQ$_1$      &     Cold Classical                   &  6.618$\pm$0.117   & 1.738$\pm$0.120  & &17, 18, 19\\ 
(19255) 1994~VK$_8$      &     Cold Classical                   &  7.016$\pm$0.163   & 1.680$\pm$0.067  & &9 \\ 
(19299) 1996~SZ$_4$      &     Plutino                          &  8.184$\pm$0.159   & 1.299$\pm$0.102  & &7, 9, 18\\ 
(19308) 1996~TO$_{66}$   &     Resonant (19:11)                 &  4.530$\pm$0.044   & 1.056$\pm$0.210  & Water ice & 6, 7, 12, 13, 14, 20, 21, f \\        
(19521) Chaos            &     Hot Classical                    &  4.442$\pm$0.069   & 1.558$\pm$0.062  & &8, 9, 10, 22\\ 
(20000) Varuna           &     Hot Classical                    &  3.345$\pm$0.059   & 1.530$\pm$0.036  & Water ice & 8, g \\                                   
(20108) 1995~QZ$_9$      &     Plutino                          &  7.889$\pm$0.399   & 1.400$\pm$0.050  & & 9, This work\\
(24835) 1995~SM$_{55}$   &     Hot Classical                    &  4.352$\pm$0.040   & 1.018$\pm$0.052  & Water ice & 8, 10, 14, 23, g\\         
(24952) 1997~QJ$_4$      &     Plutino                          &  7.389$\pm$0.114   & 1.104$\pm$0.104  & & 7, 18\\
(24978) 1998~HJ$_{151}$  &     Cold Classical                   &  7.008$\pm$0.050   & 1.820$\pm$0.042  & &19\\
(26181) 1996~GQ$_{21}$   &     Resonant (11:2)                  &  4.467$\pm$0.090   & 1.693$\pm$0.079  & Methanol & 18, 24, g\\       
(26308) 1998~SM$_{165}$  &     Resonant (2:1)                   &  5.757$\pm$0.119   & 1.620$\pm$0.105  & &9, 10, 15\\ 
(26375) 1999~DE$_9$      &     Resonant (5:2)                   &  4.810$\pm$0.046   & 1.536$\pm$0.056  & Featureless &7, 8, 25, h\\              
(28978) Ixion            &     Scattered Disk Object            &  3.366$\pm$0.038   & 1.634$\pm$0.035  & Water ice &8, h\\                
(29981) 1999~TD$_{10}$   &     Scattered Disk Object            &  8.698$\pm$0.038   & 1.230$\pm$0.028  & Water ice &8, g\\     
(31824) Elatus           &     Centaur                          & 10.439$\pm$0.107   & 1.672$\pm$0.071  & Water ice &8, 10, 26, g\\                                
(32532) Thereus          &     Centaur                          &  9.365$\pm$0.038   & 1.190$\pm$0.032  & Water ice &25, h\\                                      
(32929) 1995~QY$_9$      &     Plutino                          &  7.489$\pm$0.126   & 1.160$\pm$0.150  & &1, 13\\ 
(33001) 1997~CU$_{29}$   &     Cold Classical                   &  6.173$\pm$0.078   & 1.804$\pm$0.115  & &7, 16, 22, 27\\ 
(33128) 1998~BU$_{48}$   &     Scattered Disk Object            &  6.889$\pm$0.127   & 1.692$\pm$0.089  & &8, 10 \\
(33340) 1998~VG$_{44}$   &     Plutino                          &  6.292$\pm$0.077   & 1.511$\pm$0.055  & &8, 14, 16, 24\\ 
(35671) 1998~SN$_{165}$  &     Scattered Disk Object            &  5.431$\pm$0.068   & 1.123$\pm$0.082  & &7, 10, 16\\ 
(38083) Rhadamanthus     &     Scattered Disk Object            &  7.432$\pm$0.063   & 1.177$\pm$0.109  & &18\\ 
(38084) 1999~HB$_{12}$   &     Resonant (5:2)                   &  6.718$\pm$0.050   & 1.409$\pm$0.049  & &16, 25, 27, 28\\ 
(38628) Huya             &     Plutino                          &  4.674$\pm$0.099   & 1.539$\pm$0.062  & Featureless &29, 7, 16, 18, g\\             
(40314) 1999~KR$_{16}$   &     Scattered Disk Object            &  5.527$\pm$0.039   & 1.872$\pm$0.068  & &7, 18, 27\\ 
(42301) 2001~UR$_{163}$  &     Resonant (9:4)                   &  3.812$\pm$0.109   & 2.190$\pm$0.130  & Featureless &15, 30, 31, g\\    
(42355) Typhon           &     Scattered Disk Object            &  7.358$\pm$0.076   & 1.292$\pm$0.071  & Water ice &25, 28, h\\    
(44594) 1999~OX$_{3}$    &     Scattered Disk Object            &  6.835$\pm$0.078   & 1.839$\pm$0.087  & Water ice &8, 9, 10, 15, 21, 30, i \\
(47171) 1999~TC$_{36}$   &     Plutino                          &  4.851$\pm$0.054   & 1.740$\pm$0.049  & Water ice &10, 16, 25, 32, h\\              
(47932) 2000~GN$_{171}$  &     Plutino                          &  5.666$\pm$0.090   & 1.559$\pm$0.066  & Featureless &18, 24, h\\        
(48639) 1995~TL$_8$      &     Detached KBO                     &  4.667$\pm$0.091   & 1.693$\pm$0.217  & &8, 10, 21\\ 
(49036) Pelion           &     Centaur                          & 10.157$\pm$0.112   & 1.248$\pm$0.096  & &9, 18\\ 
(50000) Quaoar           &     Hot Classical                    &  2.220$\pm$0.029   & 1.588$\pm$0.021  & Methane &25, 33, h\\                           
(52747) 1998~HM$_{151}$  &     Cold Classical                   &  7.417$\pm$0.100   & 1.550$\pm$0.103  & &19\\ 
(52872) Okyrhoe          &     Centaur                          & 10.775$\pm$0.078   & 1.237$\pm$0.086  & Water ice &10, 16, 32, g\\                           
(52975) Cyllarus         &     Centaur                          &  8.634$\pm$0.101   & 1.803$\pm$0.102  & &8, 10, 14, 25\\ 
(53311) Deucalion        &     Cold Classical                   &  6.662$\pm$0.060   & 2.030$\pm$0.160  & &27\\ 
(54598) Bienor           &     Centaur                          &  7.727$\pm$0.077   & 1.158$\pm$0.075  & Methanol &8, 10, 15, h\\                              
(55565) 2002~AW$_{197}$  &     Hot Classical                    &  3.156$\pm$0.059   & 1.498$\pm$0.044  & Featureless &24, 33, 34, h\\     
(55576) Amycus           &     Centaur                          &  7.789$\pm$0.042   & 1.814$\pm$0.044  & Water ic &24, 28, 33, 34, i\\                     
(55636) 2002~TX$_{300}$  &     Hot Classical                    &  3.296$\pm$0.047   & 1.010$\pm$0.028  & Water ice &25, 30, q\\               
(55637) 2002~UX$_{25}$   &     Scattered Disk Object            &  3.486$\pm$0.084   & 1.502$\pm$0.052  & Water ice &24, 31, g\\                 
(55638) 2002~VE$_{95}$   &     Plutino                          &  5.143$\pm$0.062   & 1.790$\pm$0.040  & Methanol &24, g\\                 
(58534) Logos            &     Cold Classical                   &  6.759$\pm$0.181   & 1.653$\pm$0.150  & &7, 22\\ 
(59358) 1999~CL$_{158}$  &     Scattered Disk Object            &  6.653$\pm$0.090   & 1.190$\pm$0.072  & &8 \\
(60454) 2000~CH$_{105}$  &     Cold Classical                   &  6.363$\pm$0.077   & 1.699$\pm$0.083  & &28\\ 
(60458) 2000~CM$_{114}$  &     Scattered Disk Object            &  6.954$\pm$0.044   & 1.240$\pm$0.040  & &25\\ 
(60558) Echeclus         &     Centaur                          &  9.669$\pm$0.090   & 1.376$\pm$0.072  & &18, 24\\ 
(60608) 2000~EE$_{173}$  &     Scattered Disk Object            &  8.028$\pm$0.107   & 1.164$\pm$0.032  & &18, 25 \\ 
(60620) 2000~FD$_{8}$    &     Resonant (7:4)                   &  6.344$\pm$0.061   & 1.806$\pm$0.113  & &18, 28\\ 
(60621) 2000~FE$_{8}$    &     Resonant (5:2)                   &  6.510$\pm$0.062   & 1.230$\pm$0.027  & &8, 25\\ 
(63252) 2001~BL$_{41}$   &     Centaur                          & 11.273$\pm$0.065   & 1.199$\pm$0.045  & &25, 28\\ 
(65489) Ceto             &     Scattered Disk Object            &  6.205$\pm$0.060   & 1.420$\pm$0.040  & Water ice &25, g\\              
(66452) 1999~OF$_{4}$    &     Cold Classical                   &  6.255$\pm$0.090   & 1.830$\pm$0.095  & &28\\ 
(66652) Borasisi         &     Cold Classical                   &  5.420$\pm$0.051   & 1.610$\pm$0.050  & &16, 35\\ 
(69986) 1998~WW$_{24}$   &     Plutino                          &  7.964$\pm$0.096   & 1.235$\pm$0.152  & &8, 28\\ 
(69988) 1998~WA$_{31}$   &     Resonant (5:2)                   &  7.303$\pm$0.149   & 1.412$\pm$0.127  & &28\\ 
(69990) 1998~WU$_{31}$   &     Plutino                          &  7.988$\pm$0.200   & 1.225$\pm$0.086  & &28\\ 
(73480) 2002~PN$_{34}$   &     Scattered Disk Object            &  8.487$\pm$0.046   & 1.280$\pm$0.020  & Water ice &25, j\\                  
(79360) 1997~CS$_{29}$   &     Cold Classical                   &  5.068$\pm$0.085   & 1.746$\pm$0.077  & Featureless &6, 7, 14, 22, k\\     
(79978) 1999~CC$_{158}$  &     Resonant (12:5)                  &  5.409$\pm$0.091   & 1.566$\pm$0.100  & &8, 10, 24\\ 
(79983) 1999~DF$_{9}$    &     Hot Classical                    &  5.797$\pm$0.110   & 1.630$\pm$0.078  & &8\\ 
(80806) 2000~CM$_{105}$  &     Cold Classical                   &  6.302$\pm$0.030   & 1.980$\pm$0.230  & &27\\ 
(82075) 2000~YW$_{134}$  &     Resonant (8:3)                   &  4.429$\pm$0.064   & 1.417$\pm$0.077  & &21, 25, 28, 30, 31\\ 
(82155) 2001~FZ$_{173}$  &     Scattered Disk Object            &  5.811$\pm$0.027   & 1.418$\pm$0.030  & &25, 28\\ 
(82158) 2001~FP$_{185}$  &     Scattered Disk Object            &  5.940$\pm$0.053   & 1.402$\pm$0.055  & &25, 30\\ 
(83982) Crantor          &     Centaur                          &  8.693$\pm$0.057   & 1.864$\pm$0.044  & Methanol &25, 28, 33, 34, h\\                        
(84522) 2002~TC$_{302}$  &     Scattered or Detached KBO        &  3.682$\pm$0.067   & 1.741$\pm$0.048  & Water ice &21, 24, 31, g\\           
(84719) 2002~VR$_{128}$  &     Plutino                          &  5.005$\pm$0.040   & 1.540$\pm$0.040  & &24\\ 
(84922) 2003~VS$_{2}$    &     Plutino                          &  3.794$\pm$0.070   & 1.520$\pm$0.030  & Water ice &24, g\\                    
(85633) 1998~KR$_{65}$   &     Cold Classical                   &  6.599$\pm$0.073   & 1.727$\pm$0.144  & &18, 19\\ 
(86047) 1999~OY$_{3}$    &     Scattered Disk Object            &  6.293$\pm$0.055   & 1.055$\pm$0.050  & &8, 9, 18\\ 
(86177) 1999~RY$_{215}$  &     Scattered Disk Object            &  6.736$\pm$0.114   & 1.151$\pm$0.183  & &16, 18\\ 
(87269) 2000~OO$_{67}$   &     Scattered Disk Object            &  9.057$\pm$0.170   & 1.702$\pm$0.092  & &21, 25\\ 
(87555) 2000~QB$_{243}$  &     Scattered Disk Object            &  8.439$\pm$0.119   & 1.088$\pm$0.094  & &15, 28\\ 
(88269) 2001~KF$_{77}$   &     Centaur                          & 10.038$\pm$0.020   & 1.810$\pm$0.040  & &25\\ 
(90377) Sedna            &     Detached KBO                     &  1.120$\pm$0.088   & 1.874$\pm$0.115  & Methane&21, 24, 36, l\\                 
(90482) Orcus            &     Scattered Disk Object            &  1.991$\pm$0.054   & 1.042$\pm$0.037  & Methane &24, 36, m\\             
(90568) 2004~GV$_{9}$    &     Hot Classical                    &  3.786$\pm$0.080   & 1.470$\pm$0.040  & Featureless &24, h\\                         
(91133) 1998~HK$_{151}$  &     Plutino                          &  6.937$\pm$0.076   & 1.240$\pm$0.064  & &8, 16\\ 
(91205) 1998~US$_{43}$   &     Plutino                          &  7.852$\pm$0.050   & 1.185$\pm$0.102  & &28\\ 
(91554) 1999~RZ$_{215}$  &     Scattered Disk Object            &  8.072$\pm$0.079   & 1.346$\pm$0.132  & &18\\ 
(95626) 2002~GZ$_{32}$   &     Centaur                          &  6.603$\pm$0.131   & 1.199$\pm$0.075  & &25, 30, 33 \\ 
(118228) 1996~TQ$_{66}$  &     Plutino                          &  7.245$\pm$0.195   & 1.881$\pm$0.144  & &6, 7\\ 
(118378) 1999~HT$_{11}$  &     Resonant (7:4)                   &  6.906$\pm$0.040   & 1.830$\pm$0.100  & &27\\ 
(118379) 1999~HC$_{12}$  &     Scattered Disk Object            &  7.611$\pm$0.170   & 1.384$\pm$0.214  & &18\\
(118702) 2000~OM$_{67}$  &     Scattered or Detached KBO        &  7.075$\pm$0.036   & 1.290$\pm$0.040  & &21\\ 
(119068) 2001~KC$_{77}$  &     Resonant (5:2)                   &  6.822$\pm$0.030   & 1.470$\pm$0.010  & &25\\ 
(119070) 2001~KP$_{77}$  &     Resonant (7:4)                   &  6.873$\pm$0.305   & 1.720$\pm$0.319  & &28, 30\\ 
(119315) 2001~SQ$_{73}$  &     Centaur                          &  8.857$\pm$0.069   & 1.130$\pm$0.020  & &25, 31\\ 
(119473) 2001~UO$_{18}$  &     Plutino                          &  7.804$\pm$0.506   & 2.079$\pm$0.376  & &30\\ 
(119878) 2002~CY$_{224}$ &     Resonant (12:5)                  &  5.871$\pm$0.056   & 1.680$\pm$0.100  & &31\\ 
(119951) 2002~KX$_{14}$  &     Scattered Disk Object            &  4.349$\pm$0.124   & 1.660$\pm$0.040  & Featureless &24, 37, h\\                
(120061) 2003~CO$_{1}$   &     Centaur                          &  9.134$\pm$0.140   & 1.240$\pm$0.040  & &25, 27\\ 
(120132) 2003~FY$_{128}$ &     Scattered Disk Object            &  4.486$\pm$0.053   & 1.650$\pm$0.020  & Water ice &21,g \\
(120181) 2003~UR$_{292}$ &     Scattered Disk Object            &  7.093$\pm$0.100   & 1.690$\pm$0.080  & &24\\ 
(120216) 2004~EW$_{95}$  &     Plutino                          &  6.309$\pm$0.050   & 1.080$\pm$0.030  & &24\\ 
(121725) 1999~XX$_{143}$ &     Centaur                          &  8.586$\pm$0.096   & 1.734$\pm$0.145  & &8, 28\\ 
(126619) 2002~CX$_{154}$ &     Scattered or Detached KBO        &  7.178$\pm$0.075   & 1.470$\pm$0.128  & &31\\ 
(127546) 2002~XU$_{93}$  &     Scattered Disk Object            &  7.942$\pm$0.019   & 1.200$\pm$0.020  & &21\\ 
(129772) 1999~HR$_{11}$  &     Resonant (7:4)                   &  7.172$\pm$0.150   & 1.450$\pm$0.156  & &16\\ 
(130391) 2000~JG$_{81}$  &     Resonant (2:1)                   &  7.748$\pm$0.056   & 1.417$\pm$0.060  & &This work\\ 
(134860) 2000~OJ$_{67}$  &     Cold Classical                   &  6.001$\pm$0.120   & 1.720$\pm$0.078  & &8\\ 
(135182) 2001~QT$_{322}$ &     Scattered Disk Object            &  7.752$\pm$0.320   & 1.240$\pm$0.060  & &37\\ 
(136108) Haumea          &     Resonant(12:7)                   &  0.205$\pm$0.011   & 0.973$\pm$0.024  & Water ice &38, 39, n\\
(136120) 2003~LG$_{7}$   &     Resonant (3:1)                   &  8.322$\pm$0.049   & 1.271$\pm$0.091  & &This work\\ 
(136199) Eris            &     Scattered or Detached KBO        & -1.511$\pm$0.033   & 1.207$\pm$0.088  & Methane &24, 36, o \\    
(136204) 2003~WL$_{7}$   &     Centaur                          &  8.670$\pm$0.070   & 1.230$\pm$0.040  & &24\\ 
(136472) Makemake        &     Hot Classical                    & -0.317$\pm$0.024   & 1.332$\pm$0.029  & Methane &36,p \\                            
(137294) 1999~RE$_{215}$ &     Cold Classical                   &  6.091$\pm$0.073   & 1.700$\pm$0.148  & &18\\ 
(137295) 1999~RB$_{216}$ &     Resonant (2:1)                   &  7.668$\pm$0.096   & 1.419$\pm$0.142  & &18\\ 
(138537) 2000~OK$_{67}$  &     Cold Classical                   &  6.093$\pm$0.083   & 1.540$\pm$0.094  & &8\\ 
(144897) 2004~UX$_{10}$  &     Hot Classical                    &  4.216$\pm$0.087   & 1.530$\pm$0.020  & Methanol &37, i \\                    
(145480) 2005~TB$_{190}$ &     Detached KBO                     &  3.949$\pm$0.085   & 1.540$\pm$0.030  & &21\\ 
(148209) 2000~CR$_{105}$ &     Detached KBO                     &  6.191$\pm$0.073   & 1.273$\pm$0.068  & &21, 25\\ 
(148780) Altjira         &     Hot Classical                    &  5.885$\pm$0.320   & 1.640$\pm$0.170  & &30\\ 
(149560) 2003~QZ$_{91}$  &     Scattered Disk Object            &  8.302$\pm$0.028   & 1.305$\pm$0.048  & &This work\\ 
(168703) 2000~GP$_{183}$ &     Scattered Disk Object            &  5.795$\pm$0.061   & 1.160$\pm$0.057  & &8\\ 
(181708) 1993~FW         &     Hot Classical                    &  6.572$\pm$0.105   & 1.625$\pm$0.110  & &1, 17, 19, 22\\ 
(181855) 1998~WT$_{31}$  &     Hot Classical                    &  7.443$\pm$0.079   & 1.247$\pm$0.140  & &28, 40\\ 
(181867) 1999~CV$_{118}$ &     Resonant (7:3)?                  &  7.067$\pm$0.163   & 2.130$\pm$0.090  & &27\\ 
(181868) 1999~CG$_{119}$ &     Scattered Disk Object            &  7.004$\pm$0.040   & 1.530$\pm$0.080  & &27\\ 
(181871) 1999~CO$_{153}$ &     Cold Classical                   &  6.607$\pm$0.030   & 1.940$\pm$0.090  & &27\\ 
(181874) 1999~HW$_{11}$  &     Scattered or Detached KBO        &  6.706$\pm$0.062   & 1.323$\pm$0.043  & &21, 27\\ 
(182397) 2001~QW$_{297}$ &     Resonant (9:4)                   &  6.660$\pm$0.064   & 1.600$\pm$0.070  & &21\\ 
(182934) 2002~GJ$_{32}$  &     Hot Classical                    &  5.469$\pm$0.187   & 1.678$\pm$0.261  & &30, 31\\ 
1993~RO                  &     Plutino                          &  8.492$\pm$0.113   & 1.385$\pm$0.154  & &1, 9\\ 
1994~EV$_{3}$            &     Cold Classical                   &  7.110$\pm$0.072   & 1.732$\pm$0.167  & &1, 18, 27\\ 
1994~TA                  &     Centaur                          & 11.421$\pm$0.126   & 1.930$\pm$0.155  & &9, 7\\ 
1995~HM$_{5}$            &     Plutino                          &  7.849$\pm$0.109   & 1.010$\pm$0.192  & &6, 22\\ 
1995~WY$_{2}$            &     Cold Classical                   &  6.864$\pm$0.110   & 1.655$\pm$0.278  & &1, 7\\ 
1996~RQ$_{20}$           &     Hot Classical                    &  6.903$\pm$0.092   & 1.523$\pm$0.156  & &7, 10\\ 
1996~RR$_{20}$           &     Plutino                          &  6.622$\pm$0.143   & 1.868$\pm$0.130  & &7, 9, 18\\ 
1996~TK$_{66}$           &     Cold Classical                   &  6.190$\pm$0.116   & 1.666$\pm$0.088  & &7, 8, 9\\ 
1996~TS$_{66}$           &     Hot Classical                    &  5.947$\pm$0.130   & 1.665$\pm$0.157  & &6, 7, 12\\ 
1997~CV$_{29}$           &     Hot Classical                    &  7.154$\pm$0.030   & 1.860$\pm$0.022  & &19\\ 
1997~QH$_{4}$            &     Hot Classical                    &  6.996$\pm$0.136   & 1.731$\pm$0.168  & &7, 9, 10, 18 \\ 
1997~RT$_{5}$            &     Hot Classical                    &  7.117$\pm$0.140   & 1.549$\pm$0.162  & &18\\ 
1997~SZ$_{10}$           &     Resonant (2:1)                   &  8.100$\pm$0.104   & 1.790$\pm$0.085  & &9\\ 
1998~FS$_{144}$          &     Hot Classical                    &  6.717$\pm$0.105   & 1.516$\pm$0.057  & &19, 22\\ 
1998~HL$_{151}$          &     Hot Classical                    &  8.120$\pm$0.149   & 1.190$\pm$0.284  & &27, 40\\ 
1998~KG$_{62}$           &     Cold Classical                   &  6.125$\pm$0.110   & 1.602$\pm$0.158  & &16, 18\\ 
1998~KS$_{65}$           &     Cold Classical                   &  7.166$\pm$0.040   & 1.730$\pm$0.045  & &19\\ 
1998~UR$_{43}$           &     Plutino                          &  8.083$\pm$0.132   & 1.390$\pm$0.113  & &10\\ 
1998~WS$_{31}$           &     Plutino                          &  7.952$\pm$0.186   & 1.315$\pm$0.075  & &28\\ 
1998~WV$_{24}$           &     Cold Classical                   &  7.126$\pm$0.067   & 1.270$\pm$0.032  & &9\\ 
1998~WV$_{31}$           &     Plutino                          &  7.627$\pm$0.069   & 1.349$\pm$0.096  & &10, 28\\ 
1998~WX$_{24}$           &     Cold Classical                   &  6.241$\pm$0.099   & 1.790$\pm$0.071  & &9\\ 
1998~WZ$_{31}$           &     Plutino                          &  8.044$\pm$0.102   & 1.263$\pm$0.089  & &28\\ 
1998~XY$_{95}$           &     Scattered or Detached KBO        &  6.438$\pm$0.143   & 1.580$\pm$0.212  & &14\\ 
1999~CB$_{119}$          &     Hot Classical                    &  6.740$\pm$0.050   & 1.926$\pm$0.095  & &28\\ 
1999~CD$_{158}$          &     Resonant (7:4)                   &  4.837$\pm$0.111   & 1.384$\pm$0.116  & &8, 10, 40\\ 
1999~CF$_{119}$          &     Scattered or Detached KBO        &  6.982$\pm$0.084   & 1.424$\pm$0.072  & &27, 25\\ 
1999~CJ$_{119}$          &     Cold Classical                   &  6.695$\pm$0.210   & 2.070$\pm$0.220  & &27\\ 
1999~CM$_{119}$          &     Cold Classical                   &  7.356$\pm$0.060   & 1.780$\pm$0.170  & &27\\ 
1999~CQ$_{133}$          &     Hot Classical                    &  6.682$\pm$0.050   & 1.350$\pm$0.070  & &27\\ 
1999~CX$_{131}$          &     Resonant (5:3)                   &  6.914$\pm$0.087   & 1.637$\pm$0.118  & &28\\ 
1999~GS$_{46}$           &     Hot Classical                    &  6.230$\pm$0.020   & 1.760$\pm$0.070  & &27\\ 
1999~HS$_{11}$           &     Cold Classical                   &  6.344$\pm$0.081   & 1.845$\pm$0.099  & &16, 19, 28, 35\\ 
1999~HV$_{11}$           &     Cold Classical                   &  7.003$\pm$0.050   & 1.700$\pm$0.063  & &19\\ 
1999~JD$_{132}$          &     Hot Classical                    &  5.983$\pm$0.020   & 1.590$\pm$0.090  & &27\\ 
1999~OE$_{4}$            &     Cold Classical                   &  6.887$\pm$0.193   & 1.832$\pm$0.147  & &28\\ 
1999~OJ$_{4}$            &     Cold Classical                   &  6.899$\pm$0.060   & 1.675$\pm$0.077  & &28\\ 
1999~OM$_{4}$            &     Cold Classical                   &  7.521$\pm$0.100   & 1.739$\pm$0.170  & &18\\ 
1999~RJ$_{215}$          &     Scattered Disk Object            &  7.881$\pm$0.103   & 1.221$\pm$0.175  & &18\\ 
1999~RX$_{214}$          &     Cold Classical                   &  6.385$\pm$0.050   & 1.647$\pm$0.070  & &28\\ 
1999~RY$_{214}$          &     Hot Classical                    &  7.006$\pm$0.040   & 1.258$\pm$0.085  & &28\\ 
1999~TR$_{11}$           &     Plutino                          &  8.063$\pm$0.140   & 1.770$\pm$0.106  & &9\\ 
2000~AF$_{255}$          &     Scattered Disk Object            &  5.682$\pm$0.030   & 1.780$\pm$0.060  & &27\\ 
2000~CG$_{105}$          &     Hot Classical                    &  6.469$\pm$0.293   & 1.170$\pm$0.170  & &27, 40\\ 
2000~CJ$_{105}$          &     Hot Classical                    &  5.687$\pm$0.066   & 1.760$\pm$0.106  & &31\\ 
2000~CL$_{104}$          &     Cold Classical                   &  6.394$\pm$0.086   & 1.851$\pm$0.192  & &18\\ 
2000~CL$_{105}$          &     Cold Classical                   &  6.761$\pm$0.060   & 1.520$\pm$0.090  & &27\\ 
2000~CN$_{105}$          &     Cold Classical                   &  5.286$\pm$0.160   & 1.720$\pm$0.128  & &31\\ 
2000~CO$_{105}$          &     Hot Classical                    &  5.619$\pm$0.124   & 1.520$\pm$0.180  & &27\\ 
2000~CQ$_{105}$          &     Scattered Disk Object            &  5.996$\pm$0.054   & 1.107$\pm$0.043  & &25, 28\\ 
2000~FS$_{53}$           &     Cold Classical                   &  7.165$\pm$0.124   & 1.786$\pm$0.095  & &19, 27\\ 
2000~FZ$_{53}$           &     Centaur                          & 11.103$\pm$0.165   & 1.170$\pm$0.050  & &25\\ 
2000~KK$_{4}$            &     Hot Classical                    &  5.982$\pm$0.103   & 1.550$\pm$0.050  & &19\\ 
2000~PE$_{30}$           &     Scattered Disk Object            &  5.867$\pm$0.110   & 1.132$\pm$0.084  & &15, 16, 21\\ 
2000~YB$_{2}$            &     Scattered Disk Object            &  6.436$\pm$0.084   & 1.500$\pm$0.134  & &31\\ 
2001~FM$_{194}$          &     Scattered Disk Object            &  7.453$\pm$0.159   & 1.190$\pm$0.040  & &25\\ 
2001~HY$_{65}$           &     Hot Classical                    &  6.041$\pm$0.064   & 1.510$\pm$0.092  & &31\\ 
2001~HZ$_{58}$           &     Cold Classical                   &  6.158$\pm$0.053   & 1.640$\pm$0.085  & &31\\ 
2001~KA$_{77}$           &     Hot Classical                    &  5.050$\pm$0.089   & 1.812$\pm$0.122  & &8, 28, 30\\ 
2001~KB$_{77}$           &     Plutino                          &  7.349$\pm$0.078   & 1.390$\pm$0.130  & &24\\ 
2001~KD$_{77}$           &     Plutino                          &  5.928$\pm$0.096   & 1.763$\pm$0.060  & &8, 28\\ 
2001~KG$_{77}$           &     Scattered Disk Object            &  8.340$\pm$0.120   & 1.240$\pm$0.070  & &25\\ 
2001~KY$_{76}$           &     Plutino                          &  6.689$\pm$0.380   & 1.960$\pm$0.291  & &30\\ 
2001~QC$_{298}$          &     Hot Classical                    &  6.381$\pm$0.174   & 1.030$\pm$0.098  & &31\\ 
2001~QD$_{298}$          &     Hot Classical                    &  6.185$\pm$0.170   & 1.640$\pm$0.158  & &30\\ 
2001~QF$_{298}$          &     Plutino                          &  5.119$\pm$0.118   & 1.051$\pm$0.085  & &15, 24, 30\\ 
2001~QR$_{322}$          &     Neptune Trojan                   &  7.828$\pm$0.010   & 1.260$\pm$0.036  & &41\\ 
2001~QX$_{322}$          &     Scattered Disk Object            &  6.144$\pm$0.146   & 1.752$\pm$0.280  & &25, 31\\ 
2001~QY$_{297}$          &     Cold Classical                   &  5.151$\pm$0.231   & 1.561$\pm$0.177  & &15, 30, 35\\ 
2001~RZ$_{143}$          &     Cold Classical                   &  6.241$\pm$0.123   & 1.590$\pm$0.191  & &31\\ 
2001~XZ$_{255}$          &     Centaur                          & 10.800$\pm$0.080   & 1.910$\pm$0.070  & &25\\ 
2002~DH$_{5}$            &     Centaur                          & 10.115$\pm$0.100   & 1.054$\pm$0.075  & &28\\ 
2002~GB$_{32}$           &     Scattered Disk Object            &  7.638$\pm$0.019   & 1.390$\pm$0.020  & &21\\ 
2002~GF$_{32}$           &     Plutino                          &  5.973$\pm$0.210   & 1.765$\pm$0.134  & &30\\ 
2002~GH$_{32}$           &     Hot Classical                    &  6.098$\pm$0.201   & 1.509$\pm$0.160  & &30, 31\\ 
2002~GP$_{32}$           &     Resonant (5:2)                   &  6.580$\pm$0.162   & 1.386$\pm$0.162  & &30, 35\\ 
2002~GV$_{32}$           &     Plutino                          &  6.886$\pm$0.199   & 1.860$\pm$0.122  & &30\\ 
2002~MS$_{4}$            &     Resonant (18:11)                 &  3.333$\pm$0.040   & 1.070$\pm$0.040  & &24\\ 
2002~VT$_{130}$          &     Cold Classical                   &  5.426$\pm$0.092   & 2.010$\pm$0.233  & &31\\ 
2002~XV$_{93}$           &     Plutino                          &  4.434$\pm$0.040   & 1.090$\pm$0.030  & &24\\ 
2003~AZ$_{84}$           &     Plutino                          &  3.537$\pm$0.053   & 1.052$\pm$0.057  & Methanol &24, 31, 33, h\\
2003~FZ$_{129}$          &     Scattered or Detached KBO        &  6.983$\pm$0.038   & 1.320$\pm$0.040  & &21\\ 
2003~HB$_{57}$           &     Scattered or Detached KBO        &  7.389$\pm$0.028   & 1.310$\pm$0.030  & &21\\ 
2003~QA$_{92}$           &     Scattered Disk Object            &  6.367$\pm$0.240   & 1.670$\pm$0.020  & &37\\ 
2003~QK$_{91}$           &     Scattered or Detached KBO        &  6.966$\pm$0.036   & 1.370$\pm$0.040  & &21\\ 
2003~QQ$_{91}$           &     Scattered Disk Object            &  7.624$\pm$0.280   & 1.180$\pm$0.050  & &37\\ 
2003~QW$_{90}$           &     Hot Classical                    &  4.730$\pm$0.057   & 1.780$\pm$0.092  & &31\\ 
2003~TH$_{58}$           &     Plutino                          &  6.940$\pm$0.056   & 0.990$\pm$0.071  & &40\\ 
2003~UZ$_{117}$          &     Hot Classical                    &  4.920$\pm$0.083   & 0.990$\pm$0.050  & Water ice &24, q\\
2003~YL$_{179}$          &     Cold Classical                   &  7.482$\pm$0.300   & 1.260$\pm$0.090  & &37\\ 
2004~OJ$_{14}$           &     Scattered or Detached KBO        &  6.991$\pm$0.028   & 1.420$\pm$0.030  & &21\\ 
2004~UP$_{10}$           &     Neptune Trojan                   &  8.651$\pm$0.030   & 1.160$\pm$0.064  & &41\\ 
2004~XR$_{190}$          &     Detached KBO                     &  3.937$\pm$0.036   & 1.240$\pm$0.040  & &21\\ 
2005~CB$_{79}$           &     Hot Classical                    &  4.375$\pm$0.028   & 1.090$\pm$0.028  & Water ice &40, q\\
2005~EO$_{297}$          &     Resonant (3:1)                   &  7.221$\pm$0.047   & 1.320$\pm$0.050  & &21\\ 
2005~GE$_{187}$          &     Plutino                          &  7.192$\pm$0.097   & 1.740$\pm$0.112  & &40\\ 
2005~PU$_{21}$           &     Scattered Disk Object            &  6.091$\pm$0.019   & 1.790$\pm$0.020  & &21\\ 
2005~SD$_{278}$          &     Scattered or Detached KBO        &  5.915$\pm$0.019   & 1.530$\pm$0.020  & &21\\ 
2005~TN$_{53}$           &     Neptune Trojan                   &  9.027$\pm$0.040   & 1.290$\pm$0.106  & &41\\ 
2005~TO$_{74}$           &     Neptune Trojan                   &  8.426$\pm$0.030   & 1.340$\pm$0.078  & &41\\ 
2006~RJ$_{103}$          &     Neptune Trojan                   &  7.400$\pm$0.023   & 1.903$\pm$0.044  & &This work\\ 
2006~SQ$_{372}$          &     Scattered Disk Object            &  7.709$\pm$0.049   & 1.712$\pm$0.093  & &21, This work\\ 
2007~JJ$_{43}$           &     Hot Classical                    &  4.044$\pm$0.019   & 1.610$\pm$0.020  & &21\\ 
2007~JK$_{43}$           &     Scattered Disk Object            &  7.028$\pm$0.017   & 1.400$\pm$0.027  & &This work\\ 
2007~NC$_{7}$            &     Scattered Disk Object            &  8.068$\pm$0.018   & 1.282$\pm$0.028  & &This work\\ 
2007~RH$_{283}$          &     Centaur                          &  8.435$\pm$0.039   & 1.237$\pm$0.069  & &This work\\ 
2007~TG$_{422}$          &     Scattered Disk Object            &  6.186$\pm$0.010   & 1.390$\pm$0.040  & &21\\ 
2007~UM$_{126}$          &     Centaur                          & 10.161$\pm$0.042   & 1.080$\pm$0.096  & Water ice &This work, i \\                    
2007~VJ$_{305}$          &     Scattered Disk Object            &  6.713$\pm$0.028   & 1.440$\pm$0.030  & &21\\ 
2008~FC$_{76}$           &     Centaur                          &  9.181$\pm$0.039   & 1.756$\pm$0.024  & Methanol &This work, i \\                  
2008~KV$_{42}$           &     Scattered Disk Object            &  8.564$\pm$0.056   & 1.290$\pm$0.060  & &21\\ 
2008~OG$_{19}$           &     Scattered or Detached KBO        &  4.612$\pm$0.013   & 1.470$\pm$0.010  & &21\\
\end{longtable}
\end{center}

%%%%%%%%%%%%%%%%%%%%%%%%%%%%%%%%%%%%%%%%%%%%%%%%%%%
\end{appendix}

%%%%%%%%%%%%%%%%%%%%%%%%%%%%%%%%%%%%%%%%%%%%%%%%%%%
%%%%%%%%%%%%%%%%%%%%%%%%%%%%%%%%%%%%%%%%%%%%%%%%%%%

\end{document}